\documentclass[12pt,preprint,preprintnumbers,nofootinbib,groupedaddress,superscriptaddress,amsmath,amssymb]{revtex4}
\usepackage{graphicx}
\usepackage{dcolumn}
\usepackage{bm}
\usepackage{amssymb}
\usepackage{amsmath}
\usepackage{epsfig}    
\usepackage{color}
\usepackage{slashed}
\usepackage{hhline}

\def\be{\begin{equation}}
\def\ee{\end{equation}}
\newcommand{\bea}{\begin{eqnarray}}
\newcommand{\eea}{\end{eqnarray}}
\newcommand{\nn}{\nonumber}

\begin{document}

 \begin{flushright}  APCTP Pre2018 - 013  \end{flushright}

\title{A model with flavor-dependent gauged $U(1)_{B-L_1}\times U(1)_{B-L_{2}-L_{3}}$ symmetry}

\author{Linping Mu}
\email{ mulinping6266@163.com}
\affiliation{School of Physics and Information Engineering, Shanxi Normal University, Linfen 041004, China}

\author{Hiroshi Okada}
\email{okada.hiroshi@apctp.org}
\affiliation{Physics Division, National Center for Theoretical Sciences, Hsinchu, Taiwan 300}
\affiliation{Asia Pacific Center for Theoretical Physics, Pohang, Geyoengbuk 790-784, Republic of Korea}

\author{Chao-Qiang Geng}
\email{geng@phys.nthu.edu.tw}
\affiliation{Department of Physics, National Tsing Hua University, Hsinchu, Taiwan 300}
\affiliation{Physics Division, National Center for Theoretical Sciences, Hsinchu, Taiwan 300}
\affiliation{School of Physics and Information Engineering, Shanxi Normal University, Linfen 041004, China}

\date{\today}

\begin{abstract}
We  propose a new model with flavor-dependent gauged $U(1)_{B-L_1}\times U(1)_{B-L_{2}-L_{3}}$ symmetry
in addition to the flavor-blind one in the standard model.  The model contains  three right-handed neutrinos to cancel gauge anomalies and several Higgses to construct the measured fermion masses.
We show the generic feature of the model and explore its phenomenology. In  particular, we discuss the current  bounds
on  the extra gauge bosons from  the K and B meson mixings as well as the LEP and LHC data and focus on their contributions to the
 lepton flavor violating processes of $\ell_{i+1}\to \ell_i\gamma$ (i=1,2).
\end{abstract}
 \maketitle
\newpage

\section{Introduction}
It is known that two vector $U(1)$ gauge bosons often appear in grand unified theories (GUTs) such as $SO(10)$ gauged symmetry~\cite{Fritzsch:1974nn} when it spontaneously breaks down, 
in which a flavor-blind  gauged $U(1)_{B-L}$ can be naturally induced along with right-handed neutrinos.
On the other hand, the flavor-dependent  $U(1)$ gauged symmetries are one of the promising scenarios 
 to explain several anomalies beyond the standard model (SM), such as semi-leptonic decays involving $b\to s\ell\bar\ell$,  
 muon anomalous magnetic moment, and so on~\cite{ExtraZ}. 

In this paper, we propose a new model which contains two extra 
 flavor-dependent  gauge symmetries: $U(1)_{B-L_1}\times U(1)_{B-L_2-L_3}$ with the subscript numbers representing
 family indices besides  the flavor-blind SM one. This type of the extension of the SM is, of course, difficult to be embed 
 into a larger group as GUTs. 
But, due to the flavor dependence, there exist flavor changing processes via vector gauge bosons, resulting in 
a little different signatures from the typical gauged symmetries such as the flavor-blind $U(1)_{B-L}$  models.

This paper is organized as follows.
In Sec.~II, we first construct our model by showing its field contents and their charge assignments and then
give the concrete renormalizable Lagrangian with scalar  and vector gauge boson sectors.
After that, we discuss the phenomenologies, including
 the interaction terms, the bounds from the K and B meson mixings, the LEP~\cite{Schael:2013ita} and LHC~\cite{Aaboud:2017buh} experiments, and lepton flavor violations (LFVs). In Sec. III, we perform the numerical analysis.
 In Sec.~IV, we extend our model to explain several anomalies indicated from the  current experiments.
Finally, we  conclude in Sec.~V with some discussions.


\section{ Model setup and phenomenology}
First of all, we impose two additional $ U(1)_{B-L_1}\times U(1)_{B-L_2-L_3}$ gauge symmetries by including three right-handed neutral fermions $N_{R_{1,2,3}}$ with the subscripts representing the family indices.
The field contents of fermions (scalar bosons) are given in Table~\ref{tab:1} (\ref{tab:2}).
Then, the anomaly cancellations among $U(1)_{B-L_1}^3$, $U(1)_{B-L_1}$, $U(1)_{B-L_2-L_3}^3$, $U(1)_{B-L_2-L_3}$,
 $U(1)_{B-L_1}^2 U(1)_Y$, $U(1)_{B-L_1} U(1)_Y^2$, $U(1)_{B-L_2-L_3}^2 U(1)_Y$, $U(1)_{B-L_2-L_3} U(1)_Y^2$
are the same as the typical single flavor-independent gauged $U(1)_{B-L}$ symmetry, while
those from $U(1)_{B-L_1}\times U(1)_{B-L_2-L_3}^2$ and $U(1)_{B-L_1}^2\times U(1)_{B-L_2-L_3}$
are automatically cancelled because the two additional charge assignments are orthogonal each other. 
In Table~\ref{tab:2},
$H_1$ is expected to be the SM Higgs,
while $H_2$ is another isospin doublet scalar boson, which plays a role in providing the mixings of the 1-2 and 1-3 components in the 
CKM matrix, as we will see below.
Under these symmetries, the renormalizable Lagrangian for the quark and lepton sectors and scalar potential are  given by 
\begin{align}
-{\cal L}&=
 y_{u} \bar Q_{L_1}\tilde H_1 u_{R_1}  +y_{u_{ij}} \bar Q_{L_i}\tilde H_1 u_{R_j}  +y'_{u_{i1}} \bar Q_{L_i}\tilde H_2 u_{R_1}  \nn\\
&+y_{d}\bar Q_{L_1} H_1 d_{R_1}  +y_{d_{ij}} \bar Q_{L_i} H_1 d_{R_j}  +y'_{d_{1j}} \bar Q_{L_1} H_2 d_{R_j}\\
&+ y_{\nu} \bar L_{L_1}\tilde H_1 N_{R_1}  +y_{\nu_{ij}} \bar L_{L_i}\tilde H_1 N_{R_j}
+y_{\ell} \bar L_{L_1} H_1 e_{R_1}  +y_{\ell_{ij}} \bar L_{L_i} H_1 e_{R_j} \label{eq:dirac}\\
&+ \frac12 y_{N_{ij}}\varphi_1\bar N^C_{R_i} N_{R_j} + \frac12 y'_{N_i}\varphi_2 (\bar N^C_{R_1} N_{R_i} +\bar N^C_{R_i} N_{R_1}) 
+{\rm h.c.},\label{eq:majo}\\
V&= \frac{\mu_{H_1}^2}2 |H_1|^2 +  {\mu_{H_2}^2} |H_2|^2  + {\mu^2_{\varphi_1}}|\varphi_1|^2 
+ {\mu^2_{\varphi_{2}}} |\varphi_{2}|^2 \nn\\
&+
{\lambda_{H_1}} |H_1|^4 
+{\lambda_{H_2}} |H_2|^4 
+{\lambda_{\varphi_1}}|\varphi_1|^4
+{\lambda_{\varphi_{2}}}|\varphi_{2}|^4  
+ \lambda_{H_1H_2} |H_1|^2 |H_2|^2 +\lambda'_{H_1H_2} |H_1^\dag H_2|^2 
\nn\\&  
+ \lambda_{H_1\varphi_1} |H_1|^2|\varphi_1|^2 + \lambda_{H_1\varphi_{2}} |H_1|^2|\varphi_{2}|^2 
+ \lambda_{H_2\varphi_1} |H_2|^2|\varphi_1|^2 + \lambda_{H_2\varphi_{2}} |H_2|^2|\varphi_{2}|^2  ,
\label{eq:lag-lep}
\end{align}
respectively,
where $\tilde H \equiv (i \sigma_2) H^*$ with $\sigma_2$ being the second Pauli matrix, 
and $i$ runs over $2$ to $3$.\\

\begin{widetext}
\begin{center} 
\begin{table}[t]
\caption{Field contents of fermions 
and their charge assignments under $SU(3)_C\times SU(2)_L\times U(1)_Y\times U(1)_{B-L_1}\times U(1)_{B-L_2-L_3}$, where the subscripts
1 and $i=2,3$ correspond to the family indices.}\begin{tabular}{|c||c|c|c|c|c|c|c|c|c|c|c|c|c|c}\hline\hline  
Fermions& ~$Q_{L_1}$~ & ~$Q_{L_i}$~ & ~$u_{R_1}$~& ~$u_{R_i}$~ & ~$d_{R_1}$~ & ~$d_{R_i}$~ &~$L_{L_1}$~&~$L_{L_i}$~ 
& ~$e_{R_1}$~ & ~$e_{R_i}$~ & ~$N_{R_1}$~ & ~$N_{R_i}$~ 
\\\hline 
$SU(3)_C$ & $\bm{3}$  & $\bm{3}$  & $\bm{3}$ & $\bm{3}$  & $\bm{3}$  & $\bm{3}$  & $\bm{1}$  & $\bm{1}$  & $\bm{1}$ & $\bm{1}$  & $\bm{1}$  & $\bm{1}$\\\hline 
 $SU(2)_L$ & $\bm{2}$  & $\bm{2}$  & $\bm{1}$  & $\bm{1}$ & $\bm{1}$ & $\bm{1}$  & $\bm{2}$ & $\bm{2}$& $\bm{1}$ & $\bm{1}$ & $\bm{1}$ & $\bm{1}$    \\\hline 
$U(1)_Y$ & $\frac16$& $\frac16$ & $\frac23$ & $\frac23$  & $-\frac{1}{3}$ & $-\frac{1}{3}$ & $-\frac12$& $-\frac12$  & $-1$ & $-1$ & $0$ & $0$    \\\hline
 $U(1)_{B-L_1}$ & $\frac13$ & $0$  & $\frac13$ & $0$ & $\frac13$ & $0$ & $-1$& $0$  & $-1$  & $0$  & $-1$  & $0$   \\\hline
  $U(1)_{B-L_2-L_3}$  & $0$ & $\frac13$  & $0$ & $\frac13$& $0$ & $\frac13$ & $0$ & $-1$   & $0$  & $-1$ & $0$  & $-1$   \\\hline
\end{tabular}
\label{tab:1}
\end{table}
\end{center}
\end{widetext}

\begin{table}[t]
\caption{Field contents of scalar bosons 
and their charge assignments under  $SU(3)_C\times SU(2)_L\times U(1)_Y\times U(1)_{B-L_1}\times U(1)_{B-L_2-L_3}$,}
\centering {\fontsize{10}{12}
\begin{tabular}{|c||c|c|c|c|}\hline\hline
  Bosons  &~ $H_1$~ &~ $H_2$~ &~$\varphi_1$~&~$\varphi_2$~  \\\hline
$SU(3)_C$ & $\bm{1}$& $\bm{1}$& $\bm{1}$& $\bm{1}$ \\\hline 
$SU(2)_L$ & $\bm{2}$& $\bm{2}$ & $\bm{1}$& $\bm{1}$ \\\hline 
$U(1)_Y$ & $\frac12$ & $\frac12$ & $\bm{0}$  & $\bm{0}$ \\\hline
 $U(1)_{B-L_1}$ & $0$  & $\frac13$  & $0$     & $1$ \\\hline
  $U(1)_{B-L_2-L_3}$  & $0$ & $-\frac13$ & $2$  & $1$ \\\hline\end{tabular}%
} 
\label{tab:2}
\end{table}

\noindent \underline{\it Scalar sector}:

The scalar fields are parameterized as 
\begin{align}
&H_i =\left[\begin{array}{c}
w^+_i\\
\frac{v_i + h_i +i z_i}{\sqrt2}
\end{array}\right],\quad  
\varphi_i=
\frac{v'_{i} +\varphi_{R_i} + iz_{\varphi_i}}{\sqrt2},\ (i=1,2),
\label{component}
\end{align}
with all four CP-odd bosons $z_{1,2,\varphi_1,\varphi_2}$  massless, in which three of them are   
absorbed by vector gauged bosons $Z_{SM}$, $Z'$ and $Z''$, respectively, 
where $Z_{SM}\equiv (g_1^2+g_2^2)v/4$ with $v\equiv \sqrt{v_1^2+  v_2^2}\approx 246$ GeV and $Z'(Z'')$ arises from $U(1)_{B-L_1}(U(1)_{B-L_2-L_3})$.
\footnote{We remark that the dangerous physical Goldstone boson from $H_2$
can be  evaded  by introducing an isospin singlet boson $\varphi_3$ of (-1/3,1/3) under $U(1)_{B-L_1}\times U(1)_{B-L_2-L_3}$,
resulting in additional terms $(H_1^\dag H_2)\varphi_3$ and $\varphi^*_1\varphi_2\varphi_3^3/\Lambda$ that give the non-vanishing CP-odd mass. Here, $\Lambda$ is the cut-off scale, expected to be ${\cal O}(100)$ TeV  at most. Then, the CP-odd Higgs mass with ${\cal O}(100)$ GeV is found.
Even though $\varphi_3$ affects the vector gauge boson masses, we neglect the contribution hereafter, by assuming $v'_{\varphi_3}<<v'_{\varphi_{1,2}}$.
Note here that $\varphi_3$ does not contribute to the fermion masses.}
The feature of the singly charged bosons is same as the typical two-Higgs doublet model.
Consequently,  the mass-squared, mixing  and  eigenvalue-squared matrices  are  found as
\begin{align}
& M^2_C = \frac{\lambda_{H_1H_2}'}2
\left[\begin{array}{cc}
v_2^2 & v_1 v_2\\ 
v_1 v_2 & v_1^2  \\ 
\end{array}\right],\\
& O_C =
\left[\begin{array}{cc}
c_\beta & s_\beta  \\ 
- s_\beta &  c_\beta  \\ 
\end{array}\right],\quad \\
& D_C^2 
=
{\rm Diag}\left[0,  \frac{\lambda_{H_1H_2}' v^2}2 \right],
\end{align}
respectively,
where the above massless eigenstate is absorbed by the SM gauge boson $W^\pm$, and $c_\beta(s_\beta) = \cos \beta (\sin \beta)$
with $\tan \beta \equiv v_1/v_2$. 
As for the CP-even sector in the basis of $[h_1,h_2, \varphi_{R_1},\varphi_{R_2}]^t$, we get
a  four-by-four mass matrix squared $M^2_R$, which can be diagonalized by the mixing matrix $O_R$ as $D[H_1,H_2,H_3,H_4] \equiv O_R M^2_R O_R^T$, leading to $[h_1,h_2, \varphi_{R_1},\varphi_{R_2}]^t=O_R^T [H_1,H_2,H_3,H_4]^t$. Here, we identify $H_1\equiv h_{SM}$.

\noindent \underline{\it Fermion sector}:

The SM Dirac fermions are diagonalized by bi-unitary mixing matrices as $D_{u,d,e}=(U_{u,d,e})_L m_u (U^\dag_{u,d,e})_R$, 
and the active neutrinos are derived by an unitary mixing matrix as $D_{\nu}=U_\nu^* m_\nu U^\dag_\nu$,
while the observed mixing matrices can be defined by $V_{CKM}\equiv U^\dag_{uL}U_{dL}$,
and $V_{MSN}\equiv U_{\nu}^\dag U_{eL}$, respectively~{\cite{Bian:2017xzg}}.
{\it However, we impose $U_{uL}=1$ for simplicity.
Hence, we reduce the formula to be $V_{CKM}\equiv U_{dL}$.}
In the lepton sector, we classify the case of  $V_{MSN}\approx U_{\nu}^\dag$ or  $V_{MSN}\approx U_{eL}$ below.
Here, the neutrino mass matrix $m_\nu$ is induced via the canonical seesaw mechanism in Eqs.~(\ref{eq:dirac}) and (\ref{eq:majo}).

 \subsection{Neutral gauge boson sector}
 
 \noindent
\underline{\it $Z_{SM}$-$Z'$-$Z''$ mixing}:
Since $H_{2}$ and $\varphi_{1,2}$ have nonzero $U(1)_{B-L_1}$ and $U(1)_{B-L_2-L_3}$ charges, there are mixings among 
 $Z_{SM}$, $Z'$ and $Z''$.
The resulting mass matrix in the basis of $(Z_{SM},Z',Z'')$  is given by
\begin{align}
m_{Z_{SM},Z',Z''}^2
&= 
\left[\begin{array}{ccc}
\frac{g^2 v^2}4 & -\frac16 g'_1 g v_2^2 & \frac16 g'_2 g v_2^2  \\ 
-\frac16 g'_1 g v_2^2 & \frac19 g'^2_1 (v_2^2 + 9v'^2_2)  &   -\frac19 g'_1 g'_2 (v_2^2 - 9 v'^2_2) \\
\frac16 g'_2 g v_2^2  &    -\frac19 g'_1 g'_2 (v_2^2 - 9 v'^2_2)  & \frac19 g'^2_2 [v_2^2 + 9(4 v'^2_1+v'^2_2)]   \\ 
\end{array}\right],
\end{align}  
where $g^2\equiv g_1^2+g_2^2$, $m_{Z_{SM}}\equiv \frac{\sqrt{g_1^2+g_2^2}v}{2}\approx 91.18$ GeV, and
$g_1$, $g_2$, $g'_1$ and $g'_2$ are the gauge couplings of $U(1)_Y$, $SU(2)_L$, $U(1)_{B-L_1}$ and $U(1)_{B-L_2-L_3}$, respectively.
Here, we can identify the mass of $Z_1$ as the SM one, since we  expect $v_2<< v_{1} < v'_{1,2}$ in order to reproduce the SM fermion masses and the LEP measurement of $m_{Z_1}\sim m_{Z_{SM}}$.
This approximation is in good agreement with the current experimental data as the mass difference between $m_{Z_{SM}}$ and $m_{Z_1}$ should be less than  $O(10^{-3})$ GeV.

The other part can be reduced to be
\begin{align}
m_{Z',Z''}^2
&\sim 
\left[\begin{array}{cc}
 g'^2_1 v'^2_2 &    g'_1 g'_2 v'^2_2 \\
 g'_1 g'_2  v'^2_2 & g'^2_2 (4 v'^2_1+v'^2_2)   \\ 
\end{array}\right],
\end{align}  
which is diagonalized by the two-by-two mixing matrix $V_G$ as $V_G m_{Z',Z''}^2 V_G^T
\equiv {\rm Diag}(m^2_{Z'_1},m^2_{Z'_{2}})$ with 
 \begin{align}
m^2_{Z'_1}&=\frac12
\left[g'^2_2(4v'^2_1+v'^2_2) +g'^2_1 v'^2_2-\sqrt{g'^4_2(4v'^2_1+v'^2_2)^2 + g'^4_1 v'^4_2+2g'^2_1 g'^2_2 v'^2_2(-4 v'^2_1+v'^2_2)}\right],\\
m^2_{Z'_2} &=\frac12
\left[g'^2_2(4v'^2_1+v'^2_2) +g'^2_1 v'^2_2+\sqrt{g'^4_2(4v'^2_1+v'^2_2)^2 + g'^4_1 v'^4_2+2g'^2_1 g'^2_2 v'^2_2(-4 v'^2_1+v'^2_2)}\right],\\
V_G&= 
\left[\begin{array}{cc}
c_\theta&    s_\theta \\
-s_\theta & c_\theta  \\ 
\end{array}\right],\quad 
s_\theta= \frac{1}{\sqrt2} \sqrt{1+\frac{g'^2_2 (4v'^2_1+v'^2_2) -g'^2_1 v'^2_2 }{m_{Z'_2}^2 - m_{Z'_1}^2}}\label{eq.st}.
\end{align}  
Note here that we have to satisfy the following condition:
\begin{align}
16 g'^2_1  g'^2_2 v'^2_1 v'^2_2\le [g'^2_1 v'^2_2+g'^2_2 (4 v'^2_1+v'^2_2)]^2 , \label{eq-z'cond1}
\end{align}  
that arises from the vector boson masses to be positive real.

{
Here, we evaluate the typical scale of $v_2$ that should be suppressed by
the deviation of $m_{Z_1}$ from $m_{Z_{SM}}$ at the next leading order, $\delta m_Z\equiv |m_{Z_1}-m_{Z_{SM}}|$,
approximately given by 
\begin{align}
\delta m_Z^2
&\sim 
\frac{g^2 v_2^4}{72}\left(
\frac{|g'_1\sqrt{1-X}+g'_2\sqrt{1+X}|^2}{m^2_{Z_{SM}} - m^2_{Z'_1}}
+
\frac{|g'_1\sqrt{1+X}-g'_2\sqrt{1-X}|^2}{m^2_{Z_{SM}} - m^2_{Z'_2}}
\right),\\
X&=\frac{g'^2_2 (4v'^2_1+v'^2_2) -g'^2_1 v'^2_2 }{m_{Z'_2}^2 - m_{Z'_1}^2},
\end{align} 
where $\delta m_Z$ should satisfy $\delta m_Z\lesssim 2.1\times10^{-3}$ GeV from the electroweak precision test.
As a result, we find {\it e.g.}, $v_2\lesssim $19.5 GeV for $v'_{1,2}\sim 10^5$ GeV and $g'_{1,2}\sim 10^{-3}$.
}

\noindent
\underline{\it Interacting Lagrangian}:
The interactions in the kinetic term between the neutral vector bosons and quarks  in terms of the mass eigenstates are given by
\begin{align}
{\cal L}_q&=
-\frac13\left[(g'_1 c_\theta + g'_2 s_\theta)
\bar u \gamma^\mu u Z'_{1\mu}
+
(-g'_1 s_\theta + g'_2 c_\theta) \sum_{i=c,t}\bar u_i \gamma^\mu u_i Z'_{2\mu}\right]\nn\\
&-\frac13
\left[
\bar d_i \gamma^\mu(g'_1 c_\theta O_{dZ'} + g'_2 s_\theta O_{dZ''})_{ij} d_j Z'_{1\mu}
+
\bar d_i \gamma^\mu(-g'_1 s_\theta O_{dZ'} + g'_2 c_\theta O_{dZ''})_{ij} d_j Z'_{2\mu}\right],
\end{align}
with $d_{i,j}=(d,s,b)$, where $O_{dZ',dZ''}$ are found to be 
\begin{align}
O_{dZ'}= &  V_{CKM} {\rm diag}(1,0,0) V_{CKM}^\dag
\nonumber\\
\approx &
\left[\begin{array}{ccc}
0.95 & -0.22 &0.013+ 0.0032 i \\ 
-0.22 & 0.0509  & -0.0030-0.00075 i \\
0.013 - 0.0032 i & -0.0030+0.00075 i  & 0.00019 \\
\end{array}\right],\\
 O_{dZ''}= & V_{CKM} {\rm diag}(0,1,1) V_{CKM}^\dag
 \nonumber\\
 \approx &
\left[\begin{array}{ccc}
0.051 & 0.22 - 0.00014i 
&-0.0082-0.0033i \\ 
0.22+0.00014i & 0.95  & -0.0030-0.00075i \\
-0.0082+0.0033i & -0.0030+0.00075i  & 1.0 \\
\end{array}\right],
\end{align}
where we have  used the central values for the CKM elements in
$V_{CKM}$~\cite{Olive:2016xmw}.  
While the interactions between the neutral vector bosons and charged-leptons depend on the parameterizations of $V_{MNS}$,  
given by
\begin{align}
\label{L1}
{V_{MNS}\approx U^\dag_{\nu}}: &
~~{\cal L}_\ell^{(1)}=
\left[(g'_1 c_\theta + g'_2 s_\theta) \bar e \gamma^\mu e Z'_{1\mu}+
(-g'_1 s_\theta + g'_2 c_\theta)\sum_{i=\mu,\tau}\bar \ell_i \gamma^\mu \ell_i Z'_{2\mu}\right],
\\
\label{L2}
 {V_{MNS}\approx U_{eL}}:
&~~{\cal L}_\ell^{(2)}=
\left[\bar \ell_i \gamma^\mu(g'_1 c_\theta O_{\ell Z'} + g'_2 s_\theta O_{\ell Z''})_{ij} \ell_j Z'_{1\mu}+
\bar \ell_i \gamma^\mu(-g'_1 s_\theta O_{\ell Z'} + g'_2 c_\theta O_{\ell Z''})_{ij} \ell_j Z'_{2\mu}\right],
\end{align}
with $\ell_{i,j}=(e,\mu,\tau)$, where  $O_{\ell Z',\ell Z''}$ are derived as
\begin{align}
&O_{\ell Z'}= V_{MNS} {\rm diag}(1,0,0) V_{MNS}^\dag\approx\left[\begin{array}{ccc}
0.69 & -0.31-0.060i &0.33-0.068i \\  -0.31+ 0.060i & 0.14  & -0.14+0.060i \\0.33+0.068i & -0.14-0.060i  & 0.17 \\\end{array}\right], 
\end{align}
and
\begin{align}
& O_{\ell Z''}=  V_{MNS} {\rm diag}(0,1,1) V_{MNS}^\dag\approx\left[\begin{array}{ccc}
0.31 & 0.31+0.060i &0.33+0.068i  \\  0.31-0.060i  & 0.86  &  0.14-0.060i \\0.33-0.068i & 0.14+0.060i  & 0.83 \\\end{array}\right],
\end{align}
respectively, by taking the best fitted results  in ref.~\cite{Olive:2016xmw} for  $V_{MNS}$.

\subsection{Phenomenology}
Since $Z'_{1,2}$ interact with the SM fermions in a non-universal manner as discussed before,
the constraints are unlikely to be the same as those in the typical $U(1)_{B-L}$ models. 
Here, we will examine the bounds on the extra gauge bosons from the {K and B meson mixings as well as} the LEP data, and
discuss the lepton flavor violating processes of $\ell_{i+1}\to \ell_i\gamma$ (i=1 and 2).

 {    
\noindent \underline{\it 1. $M-\overline M$ meson mixings}  

The extra gauge bosons induce 
the neutral meson ($M$)-antimeson ($\overline M$) mixings with $M=(K^0,B_d,B_s)$,
such as $K^0-\bar K^0$, $B_d-\bar B_d$, and
$B_s-\bar B_s$, at the tree level.  
The formulas for the mass splittings are  given by~\cite{Gabbiani:1996hi}
\begin{align}
 \Delta m_M \approx &
\left[\frac{|(g'_1 c_\theta O_{dZ'} + g'_2 s_\theta O_{dZ''})_{21}|^2}{m^2_{Z'_1}}
+
\frac{|(-g'_1 s_\theta O_{dZ'} + g'_2 c_\theta O_{dZ''})_{21}|^2}{m^2_{Z'_2}}\right]
\nn\\
&\times\,m_M f^2_M\left[\frac{5}{12}-\frac14\left(\frac{m_M}{m_q+m_{q'}}\right)^2\right]\,,
\label{eq:kk}
\end{align}
for $M=(K^0,B_d,B_s)$ with $qq'=(ds,db,ds)$,
which should be less than the experimental values of 
$(3.48\times10^{-4},3.33\times10^{-2},1.17)\times10^{-11}$ GeV~\cite{Olive:2016xmw}, where 
$f_M=(156,191,200)$ MeV and $m_M=(0.498,5.280,5.367)$ GeV.
}
\\

\noindent \underline{\it 2. Bounds on $Z'_{1,2}$ from LEP and LHC}

From Eqs.~(\ref{L1}) and (\ref{L2}), we obtain the 
effective Lagrangians as
\begin{align}
{V_{MNS}\approx U_{\nu}^\dag}:
&~~{\cal L}_{eff}^{(1)}=
\frac{1}{2}\frac{G_1^2}{m_{Z'_1}^2}(\bar e\gamma^\mu e)(\bar e\gamma_\mu e)
+
\frac{G_1 (V_1^d)_{dd} }{3m_{Z'_1}^2}(\bar e\gamma^\mu e)(\bar d\gamma_\mu d)+
\frac{G_1^2}{3m_{Z'_1}^2}(\bar e\gamma^\mu e)(\bar u\gamma_\mu u)
,\nn\\
{V_{MNS}\approx U_{eL}}:
&~~{\cal L}_{eff}^{(2)}=\sum_{i=1,2}\left[
\frac{1}{2}\frac{(V_{ee}^{\ell})_i^2}{m_{Z'_i}^2}(\bar e\gamma^\mu e)(\bar e\gamma_\mu e)
+
\sum_{\ell'=\mu,\tau}\frac{(V_{ee}^{\ell})_i (V_{\ell'\ell'}^{\ell})_i }{m_{Z'_i}^2}(\bar e\gamma^\mu e)(\bar\ell'\gamma_\mu\ell')\right.\\
&~~~~~~~~~\left.+
\sum_{q'=d,s,b}
\frac{(V_{ee}^{\ell})_i (V^d_{q'q'})_i }{3m_{Z'_i}^2}(\bar e\gamma^\mu e)(\bar q'\gamma_\mu q')
+
\frac{(V_{ee}^{\ell})_i G_i}{3m_{Z'_i}^2}(\bar e\gamma^\mu e)(\bar u\gamma_\mu u)\right],
\end{align}
respectively, where $(V^{d(\ell)}_{ij})_1 \equiv (g'_1 c_\theta O_{d(\ell)Z'} + g'_2 s_\theta O_{d(\ell)Z''})$, $(V^{d(\ell)}_{ij})_2 \equiv (-g'_1 s_\theta O_{d(\ell)Z'} + g'_2 c_\theta O_{d(\ell)Z''})$,
$G_1 \equiv g'_1 c_\theta  + g'_2 s_\theta$, and $G_2 \equiv -g'_1 s_\theta + g'_2 c_\theta$.
As a results, the bounds for $Z'_{1,2}$ from the measurements of $e^+e^-\to f\bar f$  at LEP~\cite{Schael:2013ita}
and $q\bar q\to e\bar e(\mu\bar\mu)$ at LHC~\cite{Aaboud:2017buh}
are found to be
\begin{align}
{V_{MNS}\approx U_{\nu}^\dag}:& ~~\frac{(20.6{\rm TeV})^2}{8\pi} \lesssim\frac{m_{Z'_1}^2}{G_1^2},\quad
 \frac{(11.4{\rm TeV})^2}{12\pi}\lesssim\frac{m_{Z'_1}^2}{G_1 (V_1^d)_{dd} }\quad 
 {\rm for\ LEP}; \\
& ~~\frac{(37{\rm TeV})^2}{12\pi} \lesssim\frac{m_{Z'_1}^2}{G_1^2+(V_1^d)_{dd}G_1},\
 \frac{(30{\rm TeV})^2}{12\pi}\lesssim\frac{m_{Z'_1}^2}{G_2^2+(V_2^d)_{dd}G_2} \ 
 {\rm for\ LHC}, 
\end{align}
and 
\begin{align}
&{V_{MNS}\approx U_{eL}}: ~~\frac{(20.6{\rm TeV})^2}{8\pi} \lesssim\sum_{i=1,2}\frac{m_{Z'_i}^2}{(V_{ee}^{\ell})_i^2},\
\frac{(18.9{\rm TeV})^2}{4\pi} \lesssim\sum_{i=1,2}\frac{m_{Z'_i}^2}{(V_{ee}^{\ell})_i (V_{\mu\mu}^{\ell})_i},\nn\\
&~~\frac{(15.8{\rm TeV})^2}{4\pi}\lesssim\sum_{i=1,2}\frac{m_{Z'_i}^2}{(V_{ee}^{\ell})_i (V_{\tau\tau}^{\ell})_i }, \quad
\frac{(11.4{\rm TeV})^2}{12\pi}\lesssim\sum_{i=1,2}\frac{m_{Z'_i}^2}{(V_{ee}^{\ell})_i (V^d_{dd})_i },\nn\\ 
&~~\frac{(16.2{\rm TeV})^2}{12\pi}\lesssim\sum_{i=1,2}\frac{m_{Z'_i}^2}{(V_{ee}^{\ell})_i G_i}
\  {\rm for\ LEP}; \\
&~~ \frac{(37{\rm TeV})^2}{12\pi}\lesssim \sum_{i=1,2} \frac{m_{Z'_i}^2}{(V^\ell_{ee})_i[(V^d_{dd})_i+G_i]},\nn\\
&~~ \frac{(30{\rm TeV})^2}{12\pi}\lesssim \sum_{i=1,2} \frac{m_{Z'_i}^2}{(V^\ell_{\mu\mu})_i[(V^d_{dd})_i+G_i]} 
 \  {\rm for\ LHC}, 
\end{align}
where $f=e,\mu,\tau,d$ and $u$.
It is worthwhile mentioning that these neutral gauge boson searches will be carried out by experiments such as International Linear Collider (ILC)~\cite{Baer:2013cma},
and more stringent constraints should  be  obtained in the near future.
\\

\noindent \underline{\it 3. Lepton flavour violating processes}

For ${V_{MNS}\approx U_{\nu}^\dag}$, one does not need to consider the lepton flavor violations 
 from the $Z'_{1,2}$ mediations, because the charged-leptons are diagonal from the beginning. 
On the other hand, if ${V_{MNS}\approx U_{eL}}$,
the lepton flavor violating processes due to $Z'_{1,2}$ can be induced. In this case, we get that
\begin{align}
&{\rm BR}(\ell_b\to\ell_a\gamma)\approx\frac{48\pi^3\alpha_{em}C_{ba}}{(4\pi)^4 G_F^2}
\left| 
\frac{1}{8\pi^2}\sum_{k=e,\mu,\tau} \sum_{i=1,2}
(V^\ell_{\ell_a\ell_k})_i (V^{\ell\dag}_{\ell_k\ell_b})_i F_{II}\left[\frac{m_{\ell_k}^2}{m^2_{Z'_i}}\right]
\right|^2,
\\
& F_{II}(r)= \int_0^1 \frac{2 r x(1-x)^2}{r(1-x)+x},
\end{align}
where $G_F$  and  $\alpha_{em}$ are is the Fermi and  fine structure constants, respectively, while $C_{\mu e}\approx1$, $C_{\tau e}\approx0.1784$ and $C_{\tau\mu}\approx0.1736$. 
The current experimental limits are given by~\cite{TheMEG:2016wtm, Adam:2013mnn}:  
\begin{align}
\label{LFVs}
{\rm BR}(\mu\to e\gamma)\lesssim4.2\times10^{-13},\quad {\rm BR}(\tau\to e\gamma)\lesssim4.4\times10^{-8},\quad
{\rm BR}(\tau\to \mu\gamma)\lesssim3.3\times10^{-8}.
\end{align}
These constraints are imposed in the numerical analysis below.\footnote{One can consider the anomalous magnetic moment because of evading the stringent constraint of the trident production via the $Z'$ boson (flavor eigenstate)~\cite{trident}.
In our case, its value is of the order $10^{-14}$, which is much smaller than the experimental value.}

\begin{figure}[t]
\begin{center}
\includegraphics[width=70mm]{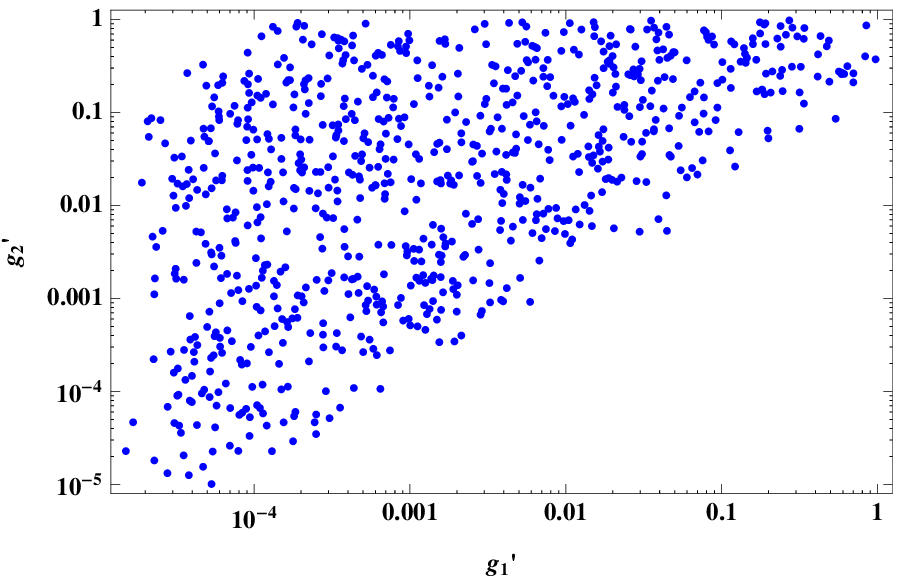} 
\includegraphics[width=70mm]{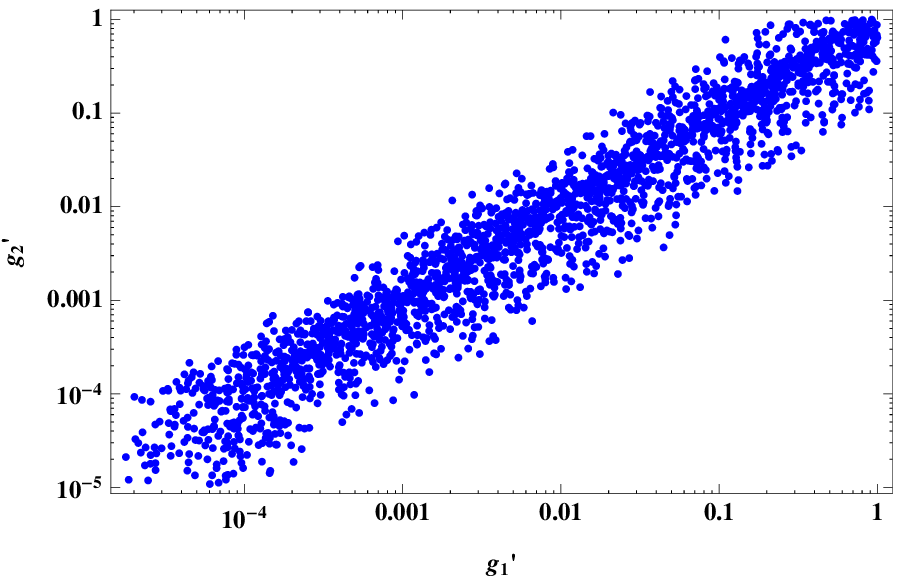} \\
\includegraphics[width=70mm]{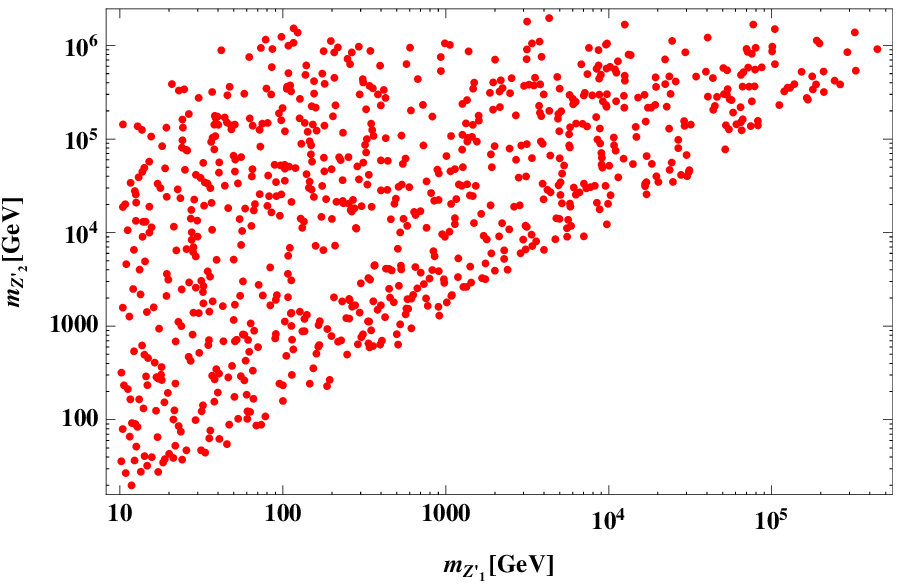} 
\includegraphics[width=70mm]{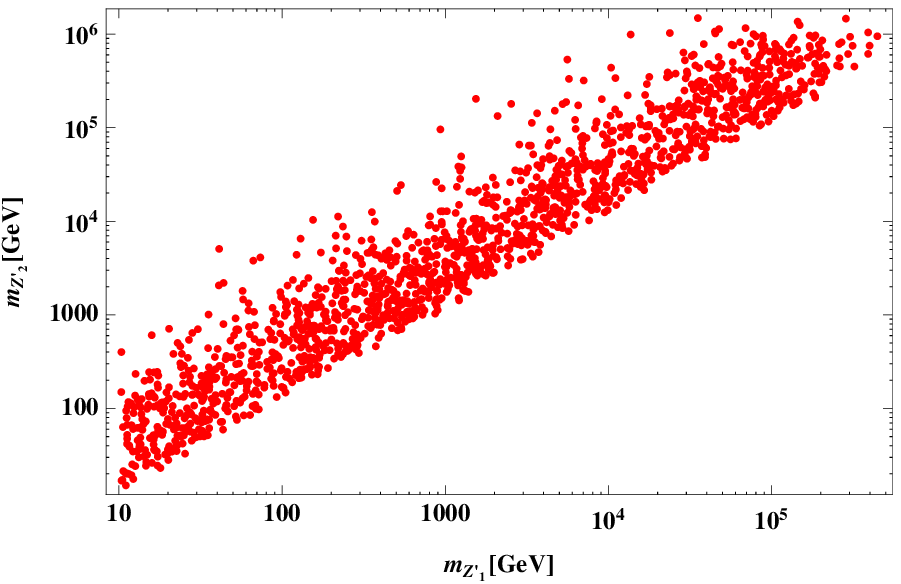} 
\caption{Allowed regions in the planes of  $g'_1$-$g'_2$ and $m_{Z'_1}$-$m_{Z'_2}$
 from the top to  bottom, 
where the left and right figures represent $s_\theta=0$ and $1/\sqrt2$ with $V_{MNS}\approx U_{\nu}^\dag$, respectively.} 
  \label{fig:Uel=1}
\end{center}\end{figure}

\section{Numerical analysis}
In our numerical analysis,
we explore the allowed  gauge parameters of $g'_{1,2}$ and $m_{Z'_{1,2}}$
by taking $s_\theta=0$ and $s_\theta=1/\sqrt2$.
We scan the parameter regions as follows: 
\begin{align}
& v'_{1,2} \in [10^3, 10^6]\ {\rm GeV}, \quad  g'_{1,2} \in [10^{-5}, 1] .
\label{eq:setting2}
\end{align}
\noindent
\underline{\it  ${V_{MNS}\approx U_{\nu}^\dag}$}:

In fig.~\ref{fig:Uel=1}, we show the allowed parameter points in the planes of  $g'_1$-$g'_2$ and $m_{Z'_1}$-$m_{Z'_2}$  from the top to bottom, 
where the left and right figures represent $s_\theta=0$ and $1/\sqrt2$, respectively.
The top-left  figure suggests a wide allowed range of values for $g'_{1,2}$  for $s_\theta=0$, 
whereas $g'_{1}$ and $g'_{2}$  should be degenerate for $s_\theta=1/\sqrt2$.  
The middle-left figure indicates that  
any values with  $m_{Z'_{1}}\le m_{Z'_{2}}$ are permitted for $s_\theta=0$, 
whereas the allowed parameter spaces for both $m_{Z'_1}$ and $m_{Z'_2}$  should be narrow within  $10\ {\rm GeV}\le m_{Z'_{1,2}}\le 10^6$ GeV  for $s_\theta=1/\sqrt2$.

\begin{figure}[t]
\begin{center}
\includegraphics[width=70mm]{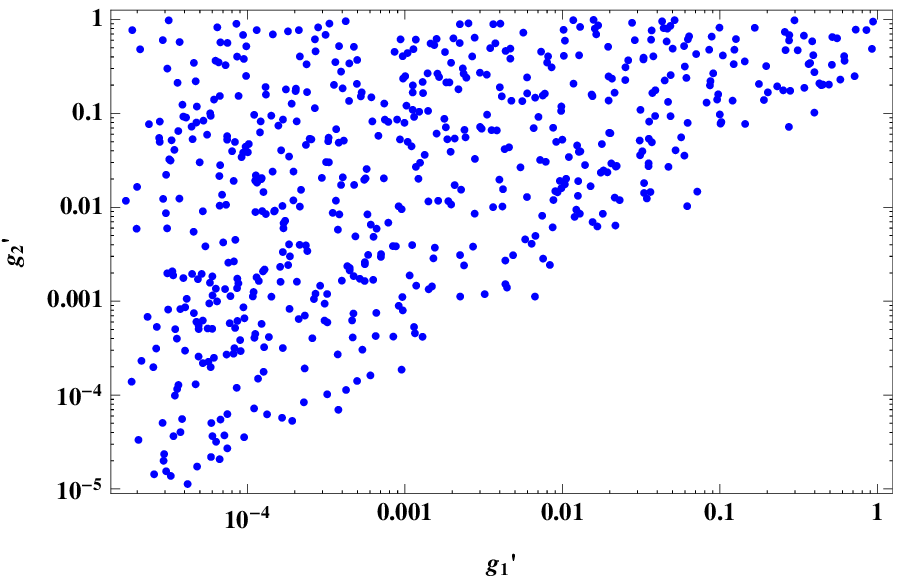} 
\includegraphics[width=70mm]{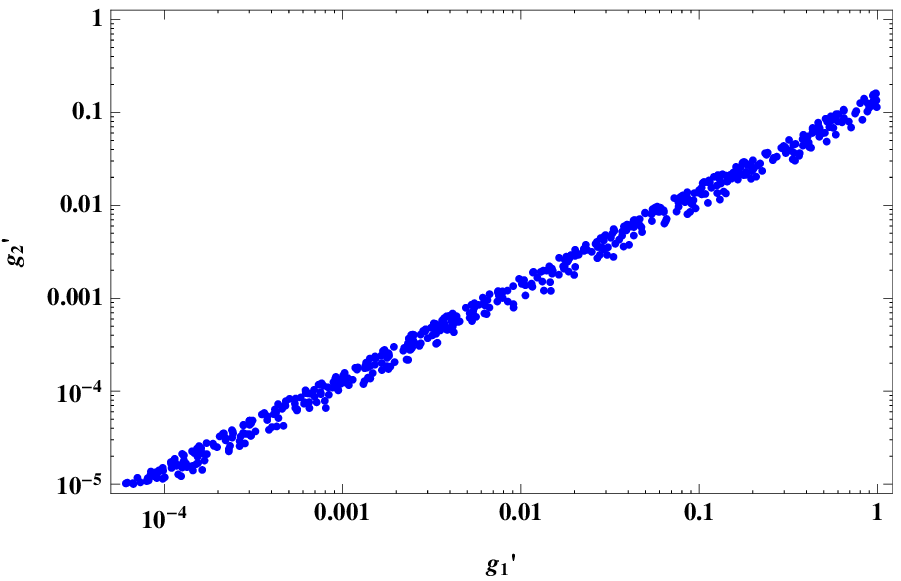} \\
\includegraphics[width=70mm]{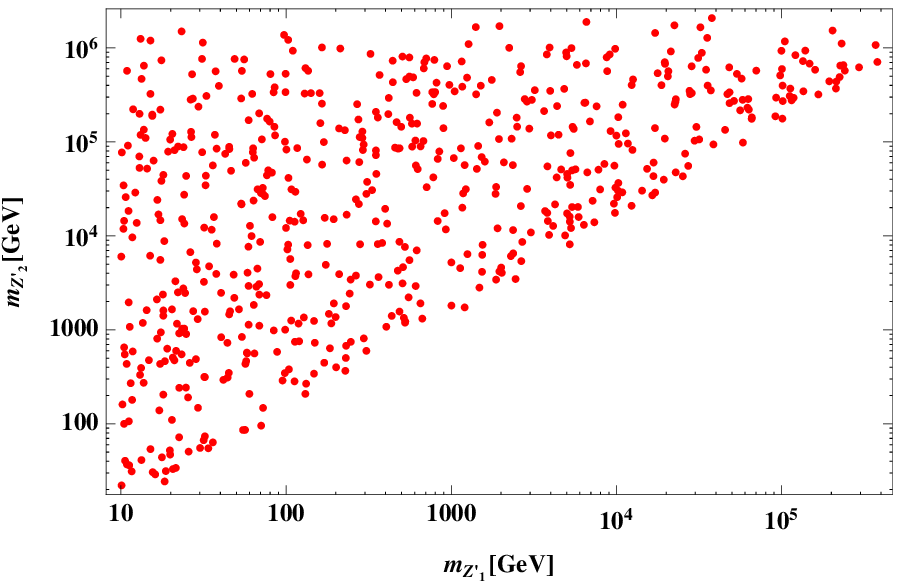} 
\includegraphics[width=70mm]{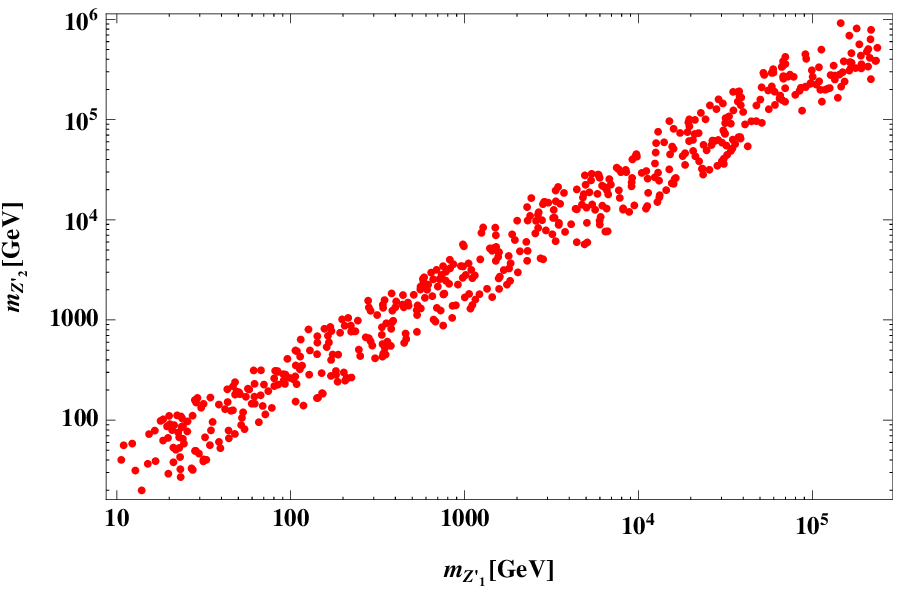} \\
\includegraphics[width=70mm]{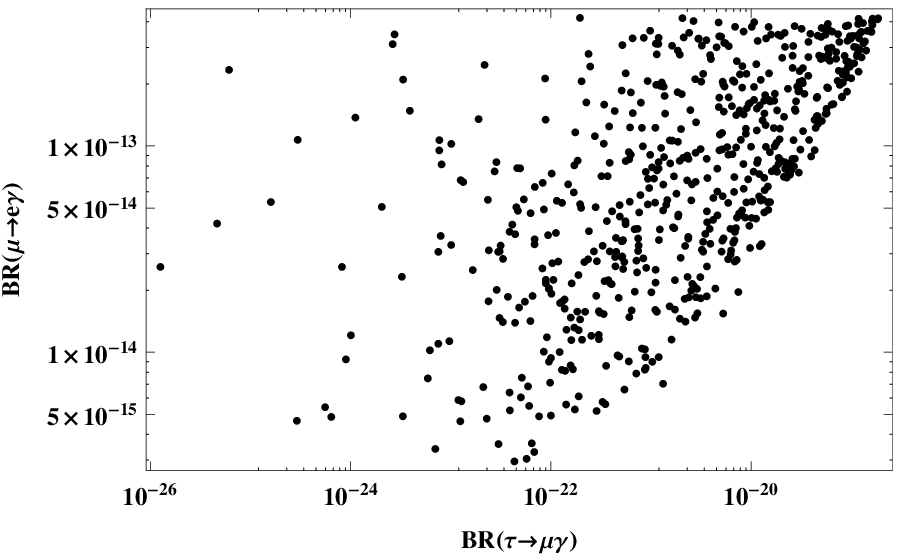} 
\includegraphics[width=70mm]{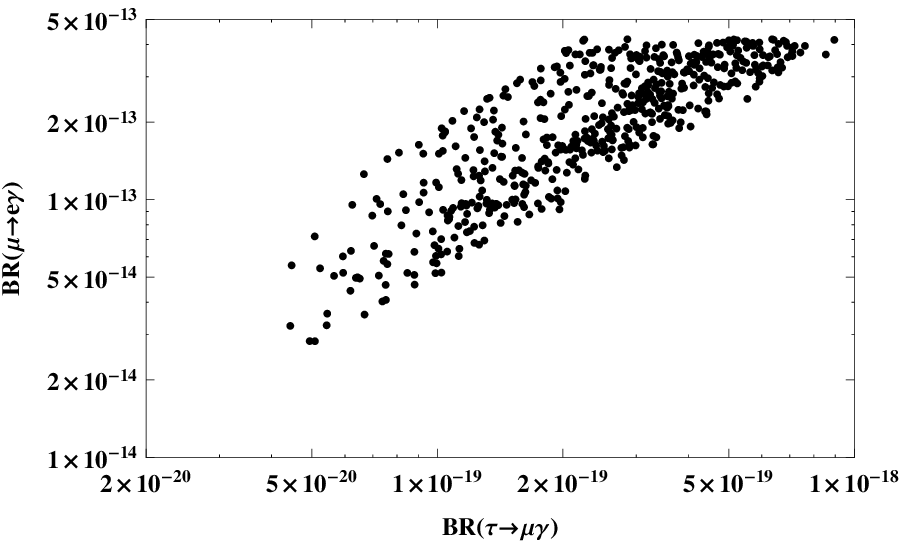} \\
\caption{Allowed regions in the planes of $g'_1$-$g'_2$, $m_{Z'_1}$-$m_{Z'_2}$,   
and ${\rm BR}(\tau\to \mu\gamma)$-${\rm BR}(\mu\to e\gamma)$ from the top to bottom, where the left and right figures represent $s_\theta=0$ and $1/\sqrt2$ with $V_{MNS}\approx U_{eL}$, respectively. } 
  \label{fig:Unu=1}
\end{center}\end{figure}
\noindent
\underline{\it ${V_{MNS}\approx U_{eL}}$}:

In fig.~\ref{fig:Unu=1}, similar to fig.~\ref{fig:Uel=1}, we illustrate the corresponding results for the case of  ${V_{MNS}\approx U_{eL}}$
by including the plots for ${\rm BR}(\tau\to \mu\gamma)$-${\rm BR}(\mu\to e\gamma)$  at the bottom.
{The generic features for the first two figures from the left-top in fig.~\ref{fig:Unu=1} 
are similar to those in fig.~\ref{fig:Uel=1}. While $g'_2$ is restricted to be $g'_2\le0.2$, the allowed regions for $g'_1$-$g'_2$ and $m_{Z'_1}$-$m_{Z'_2}$ are more degenerate than the case of $V_{MNS}\approx U_{\nu}^\dag$ for $s_\theta=1/\sqrt2$.}
For the lepton flavor violating processes at the bottom  in fig.~\ref{fig:Unu=1}, we
see  that ${\rm BR}(\mu\to e\gamma)$ reaches the current experimental bound in Eq.~(\ref{LFVs}), which is clearly
 testable in the near future  for both cases of $s_\theta$. However,
${\rm BR}(\tau\to \mu\gamma)$ is much lower than the limit in Eq.~(\ref{LFVs}).
Note here that ${\rm BR}(\tau\to e\gamma)\approx {\rm BR}(\tau\to \mu\gamma)$.
We remark that the current bounds on the masses of the extra gauge bosons are around 3 TeV
by the LHC experiments~\cite{cms-atlas}~\footnote{The LHC bounds are typically stronger than the LEP ones in case of a simple gauged $U(1)_{B-L}$ model.},  consistent with all the cases in our analyses with  $g'_1=g'_2= g_Z\approx0.72$.

Finally, we also mention that the muon anomalous magnetic moment cannot be explained in our present
model due to the constraint of the trident production via $Z'$~\cite{trident}. Moreover, the new contributions to the semi-leptonic decays of $b\to s \ell^+\ell^-$ from the $Z'_{1,2}$ mediations 
are negligible small, so that our model sheds no light to solve the recent anomalies in $B\to K^{(*)}\mu^+\mu^-$ unlike those with the extra 
$Z'$ in the literature~\cite{ExtraZ}.
Thus, we minimally extend our model to explain these issues in the next section.

\section{an extension}
We now extend our model by introducing  two extra vector-like fermions: $Q'_{L/R}= (3,2,1/6,1/2,-2/3)$ and 
$L'_{L/R}=(1,2,-1/2,1/2,-2)$ along with a neutral inert complex scalar $S=(1,1,0,-1/2,1)$ under $SU(3)_C\times SU(2)_L\times U(1)_Y\times U(1)_{B-L_1}\times U(1)_{B-L_2-L_3}$~\cite{Chiang:2017zkh}, resulting in the following additional Lagrangian:
\begin{align} 
{\cal L}=f_i \bar L_{L_i} L'_R S+ g_i \bar Q_{L_i} Q'_R S + M_{Q'} \bar Q'Q' + M_{L'} \bar L' L' + {\rm h.c.},
\end{align}
where $i=2,3$. Here, we have assumed the mass eigenstates for the above down-quark  and charged-lepton sectors
 in the SM, and $f_3<<f_2$.
As a result,  the $b\to s e\bar e$ excess is negligible that is consistent with the current experimental data, 
while $\tau\to\mu\gamma$ at one-loop level is also suppressed  to avoid the current experimental bound. 
Note that $S$ is a complex boson that is assured by the charge assignment under $U(1)_{B-L_1}\times U(1)_{B-L_2-L_3}$, and
its mass is denoted by $m_S$.

\noindent \underline{\it Muon anomalous magnetic moment \label{damu}}:

The muon anomalous magnetic moment
is formulated by the following form:
\begin{align}
\Delta a_\mu 
=
\frac{|f_{2}|^2} {{8} \pi^2} 
\int_0^1 dx \frac{x^2(1-x)}{x(x-1) + r_{L'} x + (1-x) r_S} ~,
\label{eq:damu-res}
\end{align}
where $r_{L'}\equiv (M_{L'}/m_\mu)^2$ and  $r_S \equiv (m_{S}/m_\mu)^2$.
The experimental deviation from the SM at 3.3$\sigma$ C.L. is given by~\cite{Hagiwara:2011af}
\begin{align}
\Delta a_\mu = (26.1\pm8.0(16.0))\times10^{-10} ~.
\label{eq:damu}
\end{align}

\noindent \underline{\it $B\to K^{*} \bar\mu\mu$ anomaly}:
The effective Hamiltonian for the $b \to s \mu^+ \mu^-$ transition is induced via the box diagram~\cite{Arnan:2016cpy},  given by  
\begin{align}
{\cal H}_{\rm eff}(b\to s\mu^+ \mu^-)& =
\frac{(g_{2} g^*_{3}) |f_{2}|^2 }{(4\pi)^2}  
F_{box}(m_{S},M_{Q'}, M_{L'})
(\bar s \gamma^\rho P_L b)(\bar\mu\gamma_\rho \mu - \bar\mu\gamma_\rho \gamma_5\mu)
+{\rm h.c.}
\label{eq:btosmumu}
\nn\\
%
&\equiv -C_{SM}\left[ C_9 O_9 - C_{10}  O_{10} \right]+{\rm h.c.},
\end{align}
%
where
\begin{align}
&F_{box}(m_{S},M_{Q'}, M_{L'})
\approx
\frac12\int_0^1dx_1\int_0^{1-x_1}dx_2
\frac{x_1}
{x_1 m^2_{S} +x_2 M^2_{Q'} +(1-x_1-x_2) M^2_{L'} },
\nn
\\
&C_{SM}\equiv \frac{V_{tb} V^*_{ts}G_F\alpha_{em}}{\sqrt2\pi},
\nn
\end{align}
with $V_{tb}\sim0.9991$ and $V_{ts}\sim-0.0403$ being the CKM matrix elements~\cite{Olive:2016xmw}.
Here, we take $C_9 = -C_{10}$, which is one of the promising relations to explain the anomaly~\cite{Descotes-Genon:2015uva},
and the experimental result is given by
\begin{align} 
 [-0.85,-0.50]\  {\rm at} \ 1{\sigma},\quad
 [-1.22,-0.18]\ {\rm at} \ 2{\sigma}, \label{eq:bsmm}
 \end{align}
 where  the best fit value is $-0.68$.

\noindent \underline{\it Neutral meson mixing}:
The neutral meson mixing gives the bounds on $g_i$ and $M_{Q'}$ at the low energy,
where our valid process is  the $B_s-\bar B_s$ mixing in our case.
Similar to the $B\to K^*\mu\bar\mu$ anomaly, the formula is derived by~\cite{Gabbiani:1996hi}:
\begin{align}
& \Delta m_{B_s} :
(g_{3 } g^{*}_{2})(g_{2} g^{*}_{3})  F_{box}(m_{S},M_{Q'}, M_{Q'}) \lesssim 1.17\times10^{-11}
{\times \frac{24\pi^2}{m_{B_s} f_{B_s}^2}} {{\rm GeV}}
~,
\end{align}
where the above parameters are found to be
$f_{B_s} = 0.200$~GeV~\cite{Gabbiani:1996hi}, $m_{B_s} = 5.367$~GeV~\cite{Olive:2016xmw}. 

\noindent \underline{\it  Dark matter candidate}:
We suppose that  $S$ is a DM candidate.
First, we assume that any annihilation modes coming from the Higgs potential are negligibly small.
This is a reasonable assumption, because we can avoid the strong constraint coming from the spin independent scattering cross section reported by several direct DM detection experiments, such as LUX~\cite{Akerib:2016vxi}. 
Second, we do not consider the modes through $Z'_{1,2}$ coming from the kinetic term, since this is enough suppressed by the masses of $m_{Z'_{1,2}}$. We comment here that there are two resonant solutions at around the points of $m_{Z'_1}= 2 m_S$ and $m_{Z'_2}= 2m_S$.
Subsequently, the dominant contribution to the thermal relic density comes from $f$ and $g$, and
the cross section is approximately given by~{\cite{Giacchino:2013bta}}
\begin{align}
(\sigma v _{\rm rel}) &\approx 
\frac{m_S^2}{16\pi} 
\left( \frac{ |f_{2}|^4}{6 (m_S^2+M^2_{L'})^2}
+
\sum_{i=2,3} \frac{|g_i g_j|^2}{(m_S^2+M^2_{Q'})^2} \right) v_{\rm rel}^2
+{\cal O}(v^4_{\rm rel}),
\label{eq:anni-cplx}
\end{align}
in the limit of massless final-state leptons and $m_{Q_{i,j}}^2/M^2_{Q'}<<1$. 
Here, the approximate formula is obtained by expanding the cross section in powers of the relative velocity; $v_{\rm rel}$: $\sigma v _{\rm rel}\approx a_{\rm eff} + b_{\rm eff} v^2_{\rm rel}$, where $a_{\rm eff}=0$.
The resulting relic density is found to be
\begin{align}
&\Omega h^2\approx 
\frac{1.07\times 10^9 x_f^2} {3\sqrt{g_*(x_f)} M_{\rm PL} b_{\rm eff}} 
  \label{eq:relic-cplx},
\end{align}
where the present relic density is $0.1199 \pm 0.0108$~\cite{Ade:2013zuv}, $g_*(x_f\approx 25)\approx100$ counts the degrees of freedom for relativistic 
particles, and $M_{\rm PL}\approx 1.22\times 10^{19}$~GeV is the Planck mass.

\begin{figure}[t]
\begin{center}
\includegraphics[width=70mm]{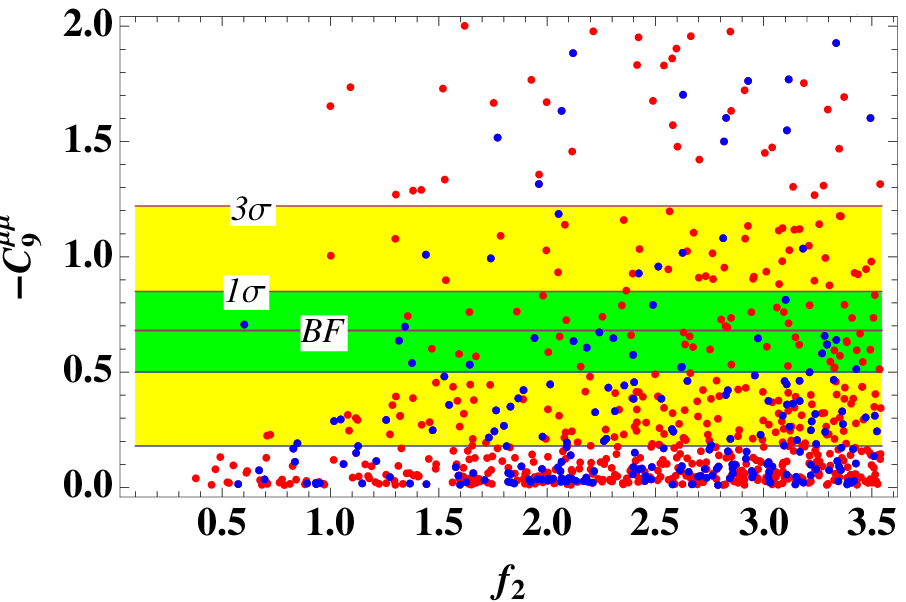} 
\includegraphics[width=70mm]{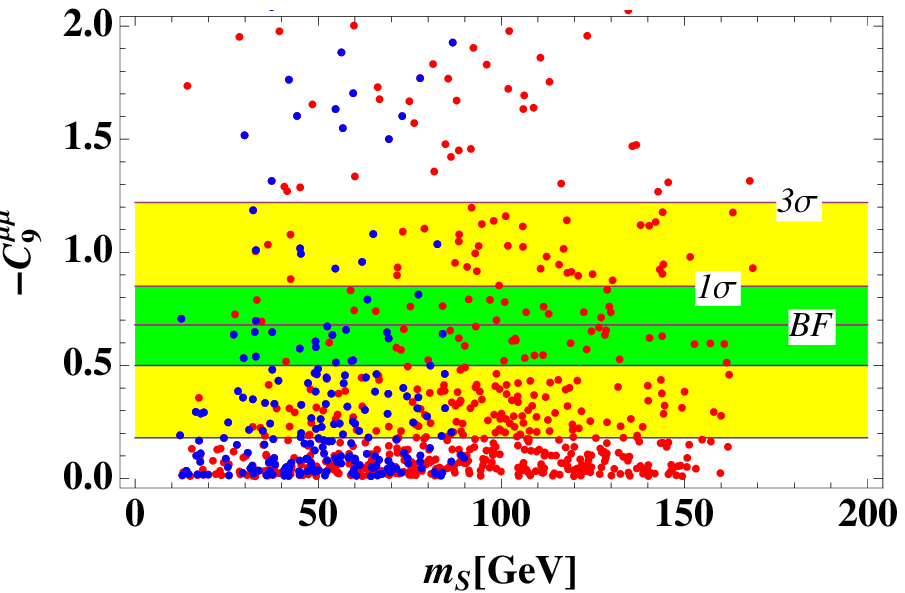} \\
\caption{Allowed regions, where the left(right)-side figure represents the $f_2(m_S)--C_{9}$ plane,
and the blue(red) points satisfies muon $g-2$ in the range of $(26.1\pm 8.0(16.0))\times10^{-10}$.
The yellow(green) region denotes the experimental allowed region [-1.22(-0.85),-0.18(-0.50)] at 1(2)$\sigma$,
where the black horizontal line inside the green region shows the best fit BF.} 
  \label{fig:c9}
\end{center}\end{figure}

\noindent \underline{\it  Numerical analyses}:

We now perform the numerical analysis to satisfy the anomalies of  the muon $g-2$, $B\to K^*\bar\mu\mu$, the constraints of 
the correct relic density, and
the neutral meson mixing, as discussed above.
We randomly select the input parameters as follows: 
\begin{align}
& f_2=[-1,\sqrt{4\pi}],\quad g_{2,3}=\pm[0.01,0],\nn\\ 
& m_S=(10,1000)\ {\rm TeV},\quad (M_{Q'},M_{L'})=(1.2 m_s,5000)\ {\rm GeV},
\label{num-2}
\end{align}
where $1.2m_S$ is used to avoid the coannihilation processes among $Q',L'$ and $S$, for simplicity. 
We show the allowed regions in fig.~\ref{fig:c9}, where the left(right)-side figure represents the $f_2(m_S)--C_{9}$ plane,
and the blue(red) points satisfies the muon $g-2$ in the range of $(26.1\pm 8.0(16.0))\times10^{-10}$ in Eq.~(\ref{eq:damu}).
The yellow(green) region denotes the experimental allowed region [-1.22(-0.85),-0.18(-0.50)] at 1(2)$\sigma$ in Eq.~(\ref{eq:bsmm}),
where the black horizontal line inside the green region corresponds to the best fit value(BF).
The left figure suggests that $f_2$ is restricted to be $[0.5,\sqrt{4\pi}]$ for both of blue and red points.
The right one implies that $m_S$ is limited to be $[10,170(90)]$ GeV in red(blue) points.

\section{Conclusions and Discussions}
We have proposed a new model with two flavor-dependent  gauge symmetries: $U(1)_{B-L_1}$ and $ U(1)_{B-L_2-L_3}$,
in addition to the SM one, along with introducing three right-handed neutrinos to cancel gauge anomalies and several scalars to construct the measured fermion masses.
We have examined the experimental bounds on
the extra gauge bosons by considering the K and B meson mixings as well as the LEP and LHC  experiments. The allowed parameter spaces for the  masses and couplings
of $Z'_{1,2}$ have been given.
Even though all the regions are within the current exclusion bounds ($\sim3$ TeV) at LHC~\cite{cms-atlas},
more stringent constraints or their discoveries will be found at ILC with
its sensitivity of the cut-off scale being  around 50-100 TeV, which are stronger than the LEP ones.

In addition, the possible effects on the flavor violating processes have been explored. Particularly, we have shown that
the branching ratio of $\mu \to e \gamma$ for the case of ${V_{MNS}\approx U_{eL}}$ can be large, which is testable by the future experiment.

Finally, we have discussed the possibility to explain the muon $g-2$, $B\to K^{(*)}\bar\mu\mu$, and dark matter candidate,
by introducing vector-like fermions $Q',L'$ and an inert complex boson $S$ with appropriate charge assignments under $SU(3)_C\times SU(2)_L\times U(1)_Y\times U(1)_{B-L_1}\times U(1)_{B-L_2-L_3}$.
We have also shown the allowed regions to satisfy all the anomalies and constraints, and found
$0.5\lesssim f_2\lesssim \sqrt{4\pi}$ for both of blue and red points, and $10\lesssim m_S\lesssim 170(90)$ GeV in red(blue) points. 
It is worthwhile to mention that Z boson decay modes of $Z\to f_i\bar f_j$ at one-loop level could restrict our parameter spaces, where $f_i$ represent all the SM fermions.
It is expected that the sensitivities of these modes further increase at future experiments, such as CEPC~\cite{CEPC-SPPCStudyGroup:2015csa}, by several orders of magnitude.

\section*{Acknowledgments}
This work was supported in part by National Center for Theoretical Sciences and
MoST (MoST-104-2112-M-007-003-MY3 and MoST-107-2119-M-007-013-MY3).
This research is supported by the Ministry of Science, ICT and Future Planning, Gyeongsangbuk-do and Pohang City (H.O.). 


\begin{thebibliography}{99}

\bibitem{Fritzsch:1974nn} 
  H.~Fritzsch and P.~Minkowski,
  Annals Phys.\  {\bf 93}, 193 (1975).
  doi:10.1016/0003-4916(75)90211-0


\bibitem{ExtraZ}
  P.~Ko, T.~Nomura and H.~Okada,
  Phys.\ Rev.\ D {\bf 95}, no. 11, 111701 (2017)
  doi:10.1103/PhysRevD.95.111701
  [arXiv:1702.02699 [hep-ph]];
  P.~Ko, T.~Nomura and H.~Okada,
  Phys.\ Lett.\ B {\bf 772}, 547 (2017)
  doi:10.1016/j.physletb.2017.07.021
  [arXiv:1701.05788 [hep-ph]];
  Y.~Tang and Y.~L.~Wu,
  Chin.\ Phys.\ C {\bf 42}, no. 3, 033104 (2018)
  doi:10.1088/1674-1137/42/3/033104
  [arXiv:1705.05643 [hep-ph]];
  C.~W.~Chiang, X.~G.~He, J.~Tandean and X.~B.~Yuan,
  Phys.\ Rev.\ D {\bf 96}, no. 11, 115022 (2017)
  doi:10.1103/PhysRevD.96.115022
  [arXiv:1706.02696 [hep-ph]];
  C.~H.~Chen and T.~Nomura,
  Phys.\ Lett.\ B {\bf 777}, 420 (2018)
  doi:10.1016/j.physletb.2017.12.062
  [arXiv:1707.03249 [hep-ph]];
  S.~Baek,
  arXiv:1707.04573 [hep-ph];
  L.~Bian, S.~M.~Choi, Y.~J.~Kang and H.~M.~Lee,
  Phys.\ Rev.\ D {\bf 96}, no. 7, 075038 (2017)
  doi:10.1103/PhysRevD.96.075038
  [arXiv:1707.04811 [hep-ph]];
  G.~Faisel and J.~Tandean,
  JHEP {\bf 1802}, 074 (2018)
  doi:10.1007/JHEP02(2018)074
  [arXiv:1710.11102 [hep-ph]];
  K.~Fuyuto, H.~L.~Li and J.~H.~Yu,
  arXiv:1712.06736 [hep-ph].
  
\bibitem{Bian:2017xzg} 
  L.~Bian, H.~M.~Lee and C.~B.~Park,
  arXiv:1711.08930 [hep-ph].
    

\bibitem{Schael:2013ita} 
  S.~Schael {\it et al.} [ALEPH and DELPHI and L3 and OPAL and LEP Electroweak Collaborations],
  Phys.\ Rept.\  {\bf 532}, 119 (2013)
  doi:10.1016/j.physrep.2013.07.004
  [arXiv:1302.3415 [hep-ex]].
  
\bibitem{Aaboud:2017buh} 
  M.~Aaboud {\it et al.} [ATLAS Collaboration],
  JHEP {\bf 1710}, 182 (2017)
  doi:10.1007/JHEP10(2017)182
  [arXiv:1707.02424 [hep-ex]].
    
\bibitem{Olive:2016xmw} 
  C.~Patrignani {\it et al.} [Particle Data Group],
  Chin.\ Phys.\ C {\bf 40}, no. 10, 100001 (2016).


\bibitem{Gabbiani:1996hi} 
  F.~Gabbiani, E.~Gabrielli, A.~Masiero and L.~Silvestrini,
  Nucl.\ Phys.\ B {\bf 477}, 321 (1996)
  doi:10.1016/0550-3213(96)00390-2
  [hep-ph/9604387].


\bibitem{Baer:2013cma} 
  H.~Baer {\it et al.},
  arXiv:1306.6352 [hep-ph].
  
\bibitem{TheMEG:2016wtm} 
  A.~M.~Baldini {\it et al.} [MEG Collaboration],
  Eur.\ Phys.\ J.\ C {\bf 76}, no. 8, 434 (2016)
  [arXiv:1605.05081 [hep-ex]].


\bibitem{Adam:2013mnn} 
  J.~Adam {\it et al.} [MEG Collaboration],
  Phys.\ Rev.\ Lett.\  {\bf 110}, 201801 (2013)
  [arXiv:1303.0754 [hep-ex]].
  
   
 \bibitem{cms-atlas}
CMS Collaboration, Physics Analysis Summary CMS PAS EXO-12-061;
G. Aad et al. [ATLAS Collaboration], ATLAS-CONF-2013-017.


       
 \bibitem{trident}
  W.~Altmannshofer, S.~Gori, M.~Pospelov and I.~Yavin,
  Phys.\ Rev.\ Lett.\  {\bf 113}, 091801 (2014)
  doi:10.1103/PhysRevLett.113.091801
  [arXiv:1406.2332 [hep-ph]].
  
  
\bibitem{Chiang:2017zkh} 
  C.~W.~Chiang and H.~Okada,
  arXiv:1711.07365 [hep-ph].


 
 
\bibitem{Hagiwara:2011af} 
  K.~Hagiwara, R.~Liao, A.~D.~Martin, D.~Nomura and T.~Teubner,
  J.\ Phys.\ G {\bf 38}, 085003 (2011)
  doi:10.1088/0954-3899/38/8/085003
  [arXiv:1105.3149 [hep-ph]].


\bibitem{Arnan:2016cpy} 
  P.~Arnan, L.~Hofer, F.~Mescia and A.~Crivellin,
  JHEP {\bf 1704}, 043 (2017)
  doi:10.1007/JHEP04(2017)043
  [arXiv:1608.07832 [hep-ph]].
  
\bibitem{Descotes-Genon:2015uva} 
  S.~Descotes-Genon, L.~Hofer, J.~Matias and J.~Virto,
  JHEP {\bf 1606}, 092 (2016)
  doi:10.1007/JHEP06(2016)092
  [arXiv:1510.04239 [hep-ph]].
  
\bibitem{Akerib:2016vxi} 
  D.~S.~Akerib {\it et al.} [LUX Collaboration],
  Phys.\ Rev.\ Lett.\  {\bf 118}, no. 2, 021303 (2017)
  doi:10.1103/PhysRevLett.118.021303
  [arXiv:1608.07648 [astro-ph.CO]].
  
\bibitem{Giacchino:2013bta} 
  F.~Giacchino, L.~Lopez-Honorez and M.~H.~G.~Tytgat,
  JCAP {\bf 1310}, 025 (2013)
  doi:10.1088/1475-7516/2013/10/025
  [arXiv:1307.6480 [hep-ph]].
  
\bibitem{Ade:2013zuv} 
  P.~A.~R.~Ade {\it et al.} [Planck Collaboration],
  Astron.\ Astrophys.\  {\bf 571}, A16 (2014)
  doi:10.1051/0004-6361/201321591
  [arXiv:1303.5076 [astro-ph.CO]].
  
    
\bibitem{CEPC-SPPCStudyGroup:2015csa} 
  CEPC-SPPC Study Group,
  IHEP-CEPC-DR-2015-01, IHEP-TH-2015-01, IHEP-EP-2015-01.
  
\end{thebibliography}
\end{document}